\documentclass[prc,aps,twocolumn,floats,floatfix,showpacs,superscriptaddress,nofootinbib]{revtex4}

\usepackage{amsmath}
\usepackage{amsfonts}
\usepackage{graphicx}
\usepackage{color}

\begin{document}

\title{Effects of the particle-particle channel on properties of
low-lying vibrational states}

\author{A. P. Severyukhin}
\affiliation{Bogoliubov Laboratory of Theoretical Physics,
             Joint Institute for Nuclear Research,
             141980 Dubna, Moscow region, Russia}

\author{V.V. Voronov}
\affiliation{Bogoliubov Laboratory of Theoretical Physics,
             Joint Institute for Nuclear Research,
             141980 Dubna, Moscow region, Russia}

\author{Nguyen Van Giai}
\affiliation{Institut de Physique Nucl\'eaire, CNRS-IN2P3,
Universit\'e Paris-Sud, F-91406 Orsay Cedex, France}

\begin{abstract}
Making use of the finite rank separable approach for the
quasiparticle random phase approximation enables one to perform
nuclear structure calculations in very large two-quasiparticle
spaces. The approach is extended to take into account the residual
particle-particle interaction. The calculations are performed by
using Skyrme interactions in the particle-hole channel and
density-dependent zero-range interactions in the particle-particle
channel. To illustrate our approach, we study the properties of
the lowest quadrupole states in the even-even nuclei $^{128}$Pd,
$^{130}$Cd, $^{124-134}$Sn, $^{128-136}$Te and $^{136}$Xe.
\end{abstract}


\date{November 15 2007}

\maketitle
%
%
\section{Introduction}
The low-energy spectrum is a key character of excitations of
nuclei in the presence of pairing correlations. The new
spectroscopic studies of exotic nuclei stimulate a development of
the nuclear models~\cite{R70,BM75,RingSchuck} to describe
properties of nuclei away from the stability line. One of the
standard tools for nuclear structure studies is the quasiparticle
random phase approximation (QRPA) with the self-consistent
mean-field derived by making use of effective nucleon-nucleon
interactions which are taken whether as non-relativistic two-body
forces~\cite{vau72,gogny} or derived from relativistic
lagrangians~\cite{r96}. Since these QRPA calculations are
performed with the same energy functional as that determining the
mean-field one does not require to introduce new parameters. Such
an approach describes the properties of the low-lying states less
accurately than more phenomenological ones, but the results are in
a reasonable agreement with experimental
data~\cite{KSGG02,PB05,vre05,ter06,rev2007}.

When the residual interaction is separable\cite{solo}, the QRPA
equations can be easily solved no matter how many
two-quasiparticle configurations are involved. Starting from an
effective interaction of the Skyrme type, a finite rank separable
approximation was proposed \cite{gsv98} for the particle-hole
(p-h) residual interaction. Such an approach allows one to perform
structure calculations in very large particle-hole spaces. Thus,
the self-consistent mean field can be calculated within the
Hartree-Fock (HF) method with the original Skyrme interactions
whereas the RPA solutions are obtained with the finite rank
approximation for the p-h matrix elements. This approach can be
extended to include the pairing correlations within the BCS
approximation~\cite{ssvg02}. Alternative schemes to factorize the
p-h interaction have also been considered in
\cite{suz81,sar99,nest02}.

Due to the anharmonicity of vibrations there is a coupling between
one-phonon and more complex states~\cite{BM75} and the complexity
of calculations beyond the standard QRPA increases rapidly with
the size of the configuration space. We have generalized our
approach to take into account a coupling between the one- and
two-phonon components of wave functions in Ref.~\cite{svg04},
where we follow the basic ideas of the quasiparticle-phonon model
(QPM)~\cite{solo}. However, the single-quasiparticle spectrum and
the parameters of the residual interaction are calculated with
Skyrme forces. Note that the QPM~\cite{solo} can achieve very
detailed predictions for nuclei away from closed
shells~\cite{gsv}, but it is difficult to extrapolate the
phenomenological parameters of the model hamiltonian to new
regions of nuclei.

In the present work, we propose an extension of our approach by
taking into account the particle-particle (p-p) residual
interaction. As an application we present the evolution of lowest
quadrupole states in even-even nuclei around the $^{132}$Sn
region. Using the neutron-rich radioactive ion beams, the recent
B(E2) measurements through Coulomb excitation in inverse
kinematics give an opportunity to compare our results and the
experimental data~\cite{Rad02,Rad05}.

This paper is organized as follows: in Sec.~II we sketch our
method, where the residual interaction is obtained by the finite
rank approximation. The hamiltonian is constructed in Sec.~IIA
whereas detailed expressions for the residual particle-particle
interaction are in Appendix A. We consider the QRPA equations in
the case of separable residual interactions in Sec.~IIB, and the
solving of these equations is explained in Appendix B. In Sec.~III
we show how this approach can be applied to treat the low-lying
states. Results of calculations for characteristics of the $2^+_1$
states in the Sn, Te isotopes and the N=82 isotones are discussed
in Sec.~IV . Conclusions are drawn in Sec.~V.
%
%
\section{Method of calculation}
\subsection{The model hamiltonian}
The starting point of the method  is the HF-BCS
calculation~\cite{RingSchuck} of the ground states. We restrict
the present discussion to the case of spherical symmetry. The
continuous part of the single-particle spectrum is discretized by
diagonalizing the HF hamiltonian on a harmonic oscillator
basis~\cite{bg77}. We work in the quasiparticle representation
defined by the canonical Bogoliubov transformation:
\begin{equation}
a_{jm}^{+} = u_j\alpha _{jm}^{+} + (-1)^{j-m}v_j\alpha _{j-m},
\end{equation}
where $jm$ denote the quantum numbers $nljm$. We use the Skyrme
interaction~\cite{sly4} in the p-h channel, while the pairing
correlations are generated by a surface peaked density-dependent
zero-range force
\begin{equation}
\label{pair}
V_{pair}({\bf r}_1,{\bf r}_2)=V_{0}\left( 1-\frac{\rho \left( r_{1}\right) }
{\rho _{c}}\right) \delta \left( {\bf r}_{1}-{\bf r}_{2}\right)
,
\end{equation}
In definition (\ref{pair}), $\rho \left( r_{1}\right)$ is the
particle density in coordinate space, $\rho _{c}$ is equal to the
nuclear saturation density, the strength $V_{0}$ is a parameter
fixed to reproduce the odd-even mass difference of nuclei in the
studied region. In Sec. III, we discuss how to make the choice of
the parameter $V_{0}$. In order to avoid divergences, it is
necessary to introduce a cut-off in the single-particle space.
This cut-off limits the active pairing space above the Fermi
level. As proposed in Ref.~\cite{softcutoff1,softcutoff2}, we have
used the smooth cut-off by multiplying the p-p matrix elements
with cut-off factors $\eta_{j}$ taken as
\begin{equation}
\eta_{j}^{2}=\left( 1+\exp (\frac{E_{j}-\lambda -\Delta E}{\mu })\right) ^{-1}
.
\end{equation}
$E_j$ are the single-particle energies, $\lambda_{n,p}$ is the
chemical potential. In our calculations we have set the energy
cut-off $\Delta E$ equal to 10 MeV above the Fermi level, the
width parameter $\mu$ being 0.5~MeV.

The residual interaction $V^{ph}_{res}$  in the p-h channel
 and $V^{pp}_{res}$ in the p-p channel can be
obtained as the second derivatives of the energy density
functional with respect to the particle density $\rho$ and the
pair density $\tilde{\rho}$, respectively. Following the method
introduced in \cite{gsv98} we  simplify $V^{ph}_{res}$ by
approximating it by its Landau-Migdal form. For Skyrme
interactions the Landau parameters are functions of the coordinate
${\bf r}$ and all parameters with $l > 1$ vanish. We keep only the
$l=0$ terms in $V^{ph}_{res}$ and the expressions for $F^{ph}_0,
G^{ph}_0, F^{'ph}_0, G^{'ph}_0$ in terms of the Skyrme force
parameters can be found in Ref.~\cite{sg81}. In this work we study
only normal parity states and one can neglect the spin-spin terms
since they play a minor role. The Coulomb and spin-orbit residual
interactions are also dropped. Therefore we can write the residual
interaction in the following form:
\begin{equation}
V^{a}_{res}({\bf r}_1,{\bf r}_2) = N_0^{-1}[ F_0^{a}(r_1)+
F_0^{'a}(r_1){\bf \tau }_1\cdot{\bf \tau }_2] \delta ({\bf r}_1 -
{\bf r}_2)
,
\end{equation}
where $a$ is the channel index $a=\{ph,pp\}$, ${\bf \tau}_i$ is
the isospin operator, and $N_0 = 2k_Fm^{*}/\pi^2\hbar^2$ with
$k_F$ and $m^{*}$ standing for the Fermi momentum and nucleon
effective mass. For the p-p channel the expressions for
$F_{0}^{pp}$ and $F_{0}^{'pp}$ have the following forms:
\begin{equation}
F_{0}^{pp}(r)=\frac{1}{4}N_{0}V_{0}\left( 1-\frac{\rho \left(
r\right) } {\rho _{c}}\right)
,
\end{equation}
\begin{equation}
F_{0}^{'pp}(r)=F_{0}^{pp}(r)
.
\end{equation}
As a matter of fact, the definition of the pairing
force~(\ref{pair}) involves the energy cut-off of the
single-particle space to restrict the active pairing space within
the mean-field approximation. This energy cut-off is still
required to eliminate the p-p matrix elements of the residual
interaction in the case of the subshells that are far from the
Fermi level. The region of influence of the residual p-p
interaction is confined to the BCS subspace
\begin{equation}
V_{1234}^{pp}={\tilde V}_{1234}^{pp}\eta_{1}\eta_{2}\eta_{3}\eta_{4}
,
\end{equation}
where all subshells below the energy cut-off are included.

The two-body matrix elements
\begin{equation}
V_{1234} = \int \phi^*_1({\bf r}_1)\phi^*_2({\bf r}_2)
V_{res}({\bf r}_1,{\bf r}_2)\phi_3({\bf r}_1)
\phi_4({\bf r}_2) {\bf dr}_1{\bf dr}_2
\end{equation}
can be written as:
\begin{widetext}
\begin{eqnarray}
V_{j_{1}m_{1}j_{2}m_{2}j_{3}m_{3}j_{4}m_{4}}^{ph} &=&{\hat J}^{-2}
\sum_{JM}\left\langle j_{1}m_{1}j_{3}-m_{3}|J-M\right\rangle \left\langle
j_{2}m_{2}j_{4}-m_{4}|JM\right\rangle (-1)^{J-M+j_{3}-m_{3}+j_{4}-m_{4}} \nonumber \\
&&\times \left\langle j_{1}\left\| i^{J}Y_{J}\right\| j_{3}\right\rangle
\left\langle j_{2}\left\| i^{J}Y_{J}\right\| j_{4}\right\rangle I^{ph}\left(
j_{1}j_{2}j_{3}j_{4}\right)
,
\end{eqnarray}
\begin{eqnarray}
\label{ppres}
V_{j_{1}m_{1}j_{2}m_{2}j_{3}m_{3}j_{4}m_{4}}^{pp} &=&\sum_{JM}\langle
j_{1}m_{1}j_{2}m_{2}\mid JM\rangle \langle j_{3}m_{3}j_{4}m_{4}\mid
JM\rangle \eta_{j_{1}}\eta_{j_{2}}\eta_{j_{3}}\eta_{j_{4}} \nonumber \\
&&\times \sum_{\lambda }\left\{
\begin{array}{ccc}
j_{4} & j_{3} & J \\
j_{1} & j_{2} & \lambda
\end{array}
\right\} (-1)^{j_{2}+j_{3}+\lambda+J}\left\langle j_{1}\left\|
i^{\lambda }Y_{\lambda }\right\| j_{3}\right\rangle \left\langle
j_{2}\left\| i^{\lambda }Y_{\lambda }\right\| j_{4}\right\rangle
I^{pp}(j_{1}j_{2}j_{3}j_{4})
\end{eqnarray}
\end{widetext}
in the p-h and p-p channels, respectively. In the above
expressions, $\hat J = \sqrt {2J+1}$, $\langle j_1 \vert\vert
i^{J}Y_{J} \vert \vert j_3 \rangle$ is the reduced matrix element
of the spherical harmonics $Y_{J \mu}$~\cite{Heyde},
$I^{a}(j_1j_2j_3j_4)$ is the radial integral:
\begin{eqnarray}
\label{I} I^{a}(j_{1}j_{2}j_{3}j_{4})
&=&N_{0}^{-1}\int_{0}^{\infty }\left( F_{0}^{a}(r)+F_{0}^{'
a}(r)\mathbf{\tau }_{1}\cdot \mathbf{\tau }
_{2}\right)  \nonumber\\
&&\times u_{j_{1}}(r)u_{j_{2}}(r)u_{j_{3}}(r)u_{j_{4}}(r)\frac{dr}{r^{2}}
,
\end{eqnarray}
where the radial wave functions $u(r)$ are related to
the single-particle wave functions:
\begin{eqnarray}
\phi_{i,m}(1)  =  \frac {u_{i}(r_1)}{r_1} {\cal Y}_{l_i,j_i}^{m}
(\hat {r_1},\sigma_1)
.
\end{eqnarray}

We see that the p-h matrix elements are in the separable form in
the angular coordinates. The separability of the antisymmetrized
p-p matrix elements is proved in Appendix~A. The radial
integrals~(\ref{I}) can be calculated accurately by choosing a
large enough cut-off radius $R$ and using a $N$-point integration
Gauss formula with abscissas ${r_k}$ and weights ${w_k}$. Thus,
the residual interaction can be reduced to a finite rank separable
form:
\begin{eqnarray}
\label{mult} \hat{V}_{res} &=&-\frac{1}{2}\sum_{a \lambda
\mu}\sum_{k=1}^{N} \sum_{\tau q=\pm 1}\left( \kappa
_{0}^{(a,k)}+q\kappa
_{1}^{(a,k)}\right)\nonumber \\
\times  &:&M_{\lambda \mu }^{\left( a,k\right) +}(\tau )M_{\lambda
\mu }^{\left( a,k\right) }(q\tau ):
.
\end{eqnarray}
We sum over the proton($p$) and neutron($n$) indexes and the notation $
\{\tau =(n,p)\}$ is used. A change $\tau \leftrightarrow -\tau $ means a
change $p\leftrightarrow n$.
$\kappa ^{(ph, k)}$ ($\kappa ^{(pp, k)}$) are the multipole interaction
strengths in the p-h (p-p) channel and they can be expressed as:
\begin{equation}
\left(
\begin{array}{c}
\kappa _{0}^{(ph,k)} \\
\kappa _{1}^{(ph,k)} \\
\kappa _{0}^{(pp,k)} \\
\kappa _{1}^{(pp,k)}
\end{array}
\right) =-N_{0}^{-1}\frac{Rw_{k}}{2r_{k}^{2}}\left(
\begin{array}{c}
F_{0}^{ph}(r_{k}) \\
F_{0}^{'ph}(r_{k})\\
\frac{1}{2}F_{0}^{pp}(r_{k}) \\
\frac{1}{2}F_{0}^{'pp}(r_{k})
\end{array}
\right)
.
\end{equation}
The multipole operators entering the normal
products in Eq.(\ref{mult}) are defined as follows:
\begin{eqnarray}
M_{\lambda \mu }^{\left( ph, k\right) +}\left( \tau \right) = {\hat \lambda }^{-1}
\left. \sum_{jj^{'}mm^{'}}\right. ^\tau (-1)^{j+m}  \nonumber \\
\times \langle jmj^{'}-m^{'}\mid \lambda \mu
\rangle f_{j'j}^{(\lambda k)}a_{jm}^{+}a_{j^{'}m^{'}}
,
\end{eqnarray}
\begin{eqnarray}
M_{\lambda \mu }^{\left( pp, k\right) +}\left( \tau \right) = \left(-1\right)^{\lambda -\mu }
{\hat \lambda }^{-1} \nonumber \\ \times \left. \sum_{jj^{^{\prime
}}mm^{'}}\right. ^\tau \langle jmj^{'}m^{'}\mid
\lambda \mu \rangle
f_{jj'}^{(\lambda k)}\eta_{j}\eta_{j'}a_{jm}^{+}a_{j^{^{\prime
}}m^{'}}^{+}
,
\end{eqnarray}
where $f_{j_1j_2}^{(\lambda k)}$ are the single-particle matrix
elements of the multipole operators:
\begin{equation}
f_{j_1j_2}^{(\lambda k)}=u_{j_1}(r_k)u_{j_2}(r_k) \langle
j_1||i^\lambda Y_\lambda ||j_2\rangle
.
\end{equation}
The residual interaction (\ref{mult}) is represented in terms of
bi-fermion quasiparticle operators and their conjugates:
\begin{equation}
B(jj^{'};\lambda \mu )\,=\,\sum_{mm^{'}}(-1)^{j^{^{\prime
}}+m{'}}\langle jmj^{'}m^{'}\mid \lambda \mu
\rangle \alpha _{jm}^{+}\alpha _{j^{'}-m^{'}}
,
\end{equation}
\begin{equation}
A^{+}(jj^{'};\lambda \mu )\,=\,\sum_{mm^{'}}\langle
jmj^{'}m^{'}\mid \lambda \mu \rangle \alpha
_{jm}^{+}\alpha _{j^{'}m^{'}}^{+}
.
\end{equation}
Thus, the hamiltonian of our method has the same form as the
hamiltonian of the QPM~\cite{solo}, but the single-quasiparticle
spectrum and the parameters of the residual interaction are
calculated with the Skyrme forces.
%
%
\subsection{QRPA equations for separable residual interactions}
We introduce the phonon creation operators
\begin{eqnarray}
Q_{\lambda \mu i}^{+}\,=\,\frac 12\sum_{jj^{'}}\left( X
_{jj^{'}}^{\lambda i}\,A^{+}(jj^{'};\lambda \mu)\right.
\nonumber\\
\left. -(-1)^{\lambda -\mu }Y _{jj^{'}}^{\lambda i}\,A(jj^{^{\prime
}};\lambda -\mu )\right)
,
\end{eqnarray}
where the index $\lambda $ denotes total angular momentum and $\mu $ is
its z-projection in the laboratory system.
One assumes that the ground state  is the QRPA phonon vacuum
$\mid 0\rangle $.
We define the excited states as $Q_{\lambda\mu i}^{+}\mid0\rangle$ with
the normalization condition
\begin{equation}
\langle 0\mid [Q_{\lambda \mu i},Q_{\lambda \mu i^{'}}^{+}] \mid
0\rangle =\delta _{ii^{'}}
\label{condition}
.
\end{equation}
Making use of the linearized equation-of-motion approach~\cite{R70}
one can get the QRPA equations~\cite{RingSchuck}:
\begin{equation}
\label{QRPA}
\left(
\begin{tabular}{ll}
$\mathcal{A}$ & $\mathcal{B}$ \\
$- \mathcal{B}$ & $- \mathcal{A}$%
\end{tabular}
\right) \left(
\begin{tabular}{l}
$ X $ \\
$ Y $
\end{tabular}
\right) =\omega \left(
\begin{tabular}{l}
$ X $ \\
$ Y $
\end{tabular}
\right)
,
\end{equation}
where the $\mathcal{A}^{(\lambda)}_{(j_1j_1^{\prime})(j_2j_2^{\prime})}$
matrix is related to forward-going graphs and the
$\mathcal{B}^{(\lambda)}_{(j_1j_1^{\prime})(j_2j_2^{\prime})}$
matrix is related to backward-going graphs.
The dimension of the matrices ${\mathcal{A}}, {\mathcal{B}}$
is the space size of the two-quasiparticle configurations.
In our case, we obtain
\begin{eqnarray}
\mathcal{A}_{(j_{1}\geq j_{1}')_{\tau }(j_{2}\geq j_{2}')
_{q\tau }}^{(\lambda )} \nonumber \\
= \varepsilon _{j_{1}j_{1}'}\delta _{j_{2}j_{1}}\delta
_{j_{2}'j_{1}'}\delta _{q1}-\hat{\lambda}^{-2}\left(
1+\delta _{j_{2}j_{2}'}\right) ^{-1} \nonumber \\
\times \sum_{k=1}^{N}f_{j_{1}j_{1}'}^{(\lambda
k)}f_{j_{2}j_{2}'}^{(\lambda k)}\left[ (\kappa
_{0}^{(ph,k)}+q\kappa _{1}^{(ph,k)})u_{j_{1}j_{1}'}^{(+)}
u_{j_{2}j_{2}'}^{(+)}\right.  \nonumber \\
\left. +(\kappa _{0}^{(pp,k)}+q\kappa _{1}^{(pp,k)})
\eta _{j_{1}j_{1}'}\eta _{j_{2}j_{2}'}\right.  \nonumber \\
\left.\times\left(
v_{j_{1}j_{1}'}^{(+)}v _{j_{2}j_{2}'}^{(+)}+
v_{j_{1}j_{1}'}^{(-)}v _{j_{2}j_{2}'}^{(-)}\right)
\right]
,
\end{eqnarray}
\begin{eqnarray}
\mathcal{B}_{(j_{1}\geq j_{1}')_{\tau }(j_{2}\geq j_{2}')
_{q\tau }}^{(\lambda )} = -\hat{\lambda}^{-2}\left( 1+\delta
_{j_{2}j_{2}'}\right) ^{-1}  \nonumber \\
\times \sum_{k=1}^{N}f_{j_{1}j_{1}'}^{(\lambda
k)}f_{j_{2}j_{2}'}^{(\lambda k)}\left[ (\kappa
_{0}^{(ph,k)}+q\kappa _{1}^{(ph,k)})u_{j_{1}j_{1}'}^{(+)}
u_{j_{2}j_{2}'}^{(+)}\right.  \nonumber \\
\left. -(\kappa _{0}^{(pp,k)}+q\kappa _{1}^{(pp,k)})
\eta _{j_{1}j_{1}'}\eta _{j_{2}j_{2}'}\right.  \nonumber \\
\left.\times\left(v_{j_{1}j_{1}'}^{(+)}v _{j_{2}j_{2}'}^{(+)}-
v_{j_{1}j_{1}'}^{(-)}v _{j_{2}j_{2}'}^{(-)}\right)
\right]
,
\end{eqnarray}
where $ \varepsilon _{jj'}=\varepsilon _j+\varepsilon _{j'}$,
$\eta_{jj'}=\eta_j+\eta_{j'}$,
$u_{jj'}^{(+)}=u_jv_{j'}+ v_ju_{j'} $ and $
v_{jj'}^{(\pm)}=u_ju_{j'}\pm v_jv_{j'} $ . The explicit solution
of the corresponding QRPA equations is given in Appendix~B. Thus,
this approach enables one to reduce remarkably the dimensions of
the matrices that must be inverted to perform structure
calculations in very large configuration spaces. It is shown that
the matrix dimensions never exceed $6N \times 6N$ independently of
the configuration space size. If we omit the residual interaction
in the p-p channel then the matrix dimension is reduced by a
factor 3 \cite{gsv98,ssvg02}.
%
%
\section{Details of calculations}
We apply our approach to study characteristics of the lowest
vibrational states in the nuclei around the $^{132}$Sn region. In
this work we use the parametrization SLy4~\cite{sly4} of the
Skyrme interaction. One peculiarity is that the parameters of the
force have been adjusted to describe the pure neutron matter. It
follows that this parametrization is a good candidate to describe
isotopic properties of nuclei from the $\beta$-stability line to
the neutron drip line. In our calculations the single-particle
continuum is discretized~\cite{bg77} by diagonalizing the HF
hamiltonian on a basis of twelve harmonic oscillator shells and
cutting off the single-particle spectra at the energy of 100 MeV.
This is sufficient to exhaust practically all the energy-weighted
sum rule within the QRPA. We use the isospin-invariant
surface-peaked pairing force~(\ref{pair}). The value $\rho_{c}$=
0.16fm~$^{-3}$ is the nuclear saturation density for the SLy4
force. The pairing strength $V_{0}$ is fitted to reproduce the
pairing energies given by
\begin{equation}
\label{P}
P_{N}=\frac{1}{2}\left( B\left( N,Z\right) +B\left( N+2,Z\right) -2B\left(
N-1,Z\right) \right)
\end{equation}
for neutrons, and similarly for protons.
The strength $V_{0}$ is taken equal to -940 MeVfm$^{3}$ in order
to get a reasonable description of the energies~(\ref{P}) for both
protons and neutrons. The Landau parameters $F^{ph}_0$,
$F^{ph'}_0$, $G^{ph}_0$, $G^{ph'}_0$ expressed in terms of the
Skyrme force parameters \cite{sg81} depend on $k_F$. As it is
pointed out in our previous works \cite{gsv98,ssvg02} one needs to
adopt some effective value for $k_F$ to give an accurate
representation of the original p-h Skyrme interaction. For the
present calculations we use the nuclear matter value for $k_F$.
Our previous investigations \cite{ssvg02,svg04} enable us to
conclude that $N$=45 for the rank of our separable approximation
is enough for multipolarities $\lambda \le 6 $ in nuclei with
$A\le 208$.

It is worth to mention the significance of the energy cut-off of
the single-particle space to confine the active space of the
residual p-p interaction. Our choice for the cut-off eliminates
matrix elements~(\ref{ppres}) coupling single-particle states
inside and outside of the BCS subspace. As it is seen from Table I
omitting the energy cut-off would lead to
an overestimation of the effect of the residual p-p interaction on
the $2^+_1$ energy and $B(E2\uparrow)$ in $^{130}$Te, for example.
\begin{table}
\caption{ Properties of the $2^+_1$ state in $^{130}$Te as an
illustrative example to demonstrate of effect of the residual p-p
interaction.}
\begin{tabular}{lcc}
\hline\noalign{\smallskip}
residual    & $E$  & $B(E2\uparrow)$\\
interaction & (MeV)& (e$^2$ fm$^4$) \\
\noalign{\smallskip}\hline\noalign{\smallskip}
 ph            & 1.49     & 3400  \\
 ph+pp         & 1.15     & 4000  \\
\noalign{\smallskip}\hline\noalign{\smallskip}
 ph+pp (cut-off)& 1.27    & 3600  \\
\noalign{\smallskip}\hline
\end{tabular}
\end{table}
\section{Results}
\subsection{ Sn isotopes}
As the first application of the method we investigate the p-p
channel effects on energies and transition probabilities of
$2^+_1$ states in $^{124-134}$Sn. Results of our calculations for
the $2^+_1$ energies and $B(E2)$ transition probabilities are
compared with experimental data \cite{Ram01,Rad02,Rad05} in Fig.1.
As it is seen from  Fig.1 there is a remarkable increase of the
$2^+_1$ energy and $B(E2\uparrow)$ in $^{132}$Sn in comparison
with those in $^{130,134}$Sn. As it was explained in our previous
paper \cite{svg04} such a behaviour of $B(E2\uparrow)$ is related
with the proportion between the QRPA amplitudes for neutrons and
protons in Sn isotopes. Including the p-p channel changes
contributions of the main configurations only slightly, but the
general structure of the $2^+_1$ remains the same. The neutron
amplitudes are dominant in all Sn isotopes and the contribution of
the main neutron configuration $\{1h_{11/2},1h_{11/2}\}$ increases
from 58\% (61\% in the case of the inclusion the p-h interaction
only) in $^{124}$Sn to 85.6\% (85.3\% for the p-h case) in
$^{130}$Sn when neutrons fill the subshell $1h_{11/2}$. At the
same time the contribution of the main proton configuration
$\{2d_{5/2},1g_{9/2}\}$ is decreasing from  15\% in $^{124}$Sn
to 7\% in $^{130}$Sn. The closure of the neutron subshell
$1h_{11/2}$ in $^{132}$Sn leads to the vanishing of the neutron
pairing. The energy of the first neutron two-quasiparticle pole
$\{2f_{7/2},1h_{11/2}\}$ in $^{132}$Sn is larger than energies of
the first poles in $^{130,134}$Sn and the contribution of the
$\{2f_{7/2},1h_{11/2}\}$ configuration in the doubly magic
$^{132}$Sn is about 61\%. Furthermore, the first pole in
$^{132}$Sn is closer to the proton poles. This means that the
contribution of the proton two-quasiparticle configurations is
larger than those in the neighbouring isotopes and as a result the
main proton configuration $\{2d_{5/2},1g_{9/2}\}$ in $^{132}$Sn
exhausts about 33\%. In $^{134}$Sn the leading contribution (about
96\%) comes from the neutron configuration $\{2f_{7/2},2f_{7/2}\}$
and as consequently the $B(E2)$ value is reduced. Such a behaviour
of the $2^+_1$ energies and $B(E2)$ values in the neutron-rich Sn
isotopes reflects the shell structure in this region. As one can
see from Fig.1 the inclusion of the p-p channel results in a
reduction of energies and transition probabilities. The
calculations reproduce very well a general behaviour for energies
and transition probabilities, but there is some overestimation in
comparison with experimental data. One can expect an improvement
if the coupling with the two-phonon components of the wave
functions \cite{svg04} is taken into account. Such calculations
are now in progress. It is worth to mention that the first
prediction of the anomalous behaviour of $2^+$ excitations around
$^{132}$Sn based on the QRPA calculations with a separable
quadrupole-plus-pairing hamiltonian has been done in \cite{Ter02}.
Other QRPA calculations with Skyrme \cite{ter06,colo03} and Gogny
\cite{giam03} forces give similar results for Sn isotopes.
\begin{figure}[t!]
\includegraphics[width=7cm]{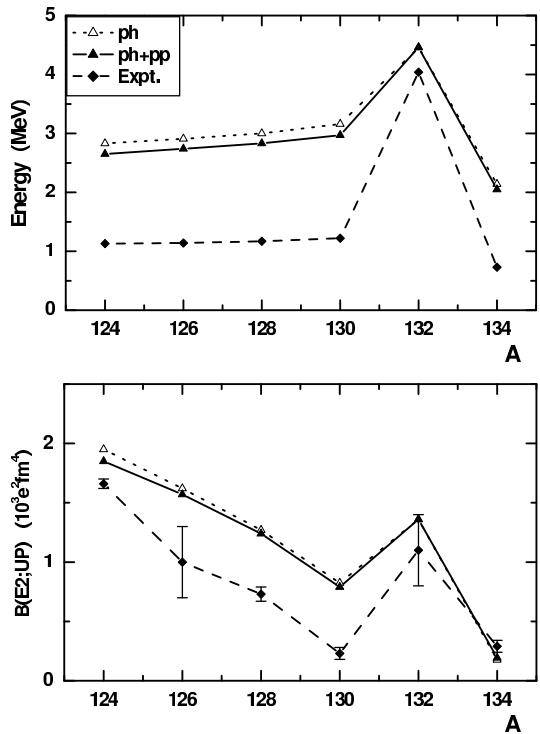}
\caption{Energies and B(E2)-values for up-transitions
to the first $2^{+}$ states in $^{124-134}$Sn.}
\end{figure}
%
%
%
\subsection{Te isotopes}
Let us now discuss the Te isotopes. The calculated $2^+_1$ state
energies and transition probabilities B(E2) in the $^{128-136}$Te
isotopes and experimental data \cite{Rad02,Rad05,Ram01} are shown
in Fig.2. The general behaviour of energies of the Te isotopes is
similar to that of the Sn isotopes. They have a maximal value at
N=82, but the behaviour of the B(E2)-values is different and
corresponds to a standard evolution of the $B(E2)$ near closed
shells. As it is seen from Fig.2 there is a decrease of
the $2^+_1$ energies due to the inclusion of the p-p channel. At
the same time the $B(E2)$-values do not change practically. It
means that the collectivity of the $2^+_1$ states is reduced.  The
neutron configurations exhaust about 17\% and 28\% of the wave
function normalization in $^{132}$Te and $^{136}$Te respectively.
In $^{134}$Te the contribution of the neutron configurations is
less than 3\% and the dominant proton configuration
$\{1g_{7/2},1g_{7/2}\}$ gives a contribution of about 65\% that
is almost twice larger than in the neighboring Te isotopes. As
far as a contribution of this configuration into the transition
probability is less than contributions of other proton
configurations the $B(E2)$-values have such a behaviour near N=82.
The structure of the $2^+_1$ in $^{132}$Te is similar to that in
$^{136}$Te and as a result the $B(E2)$-values differ slightly.

Our calculations describe correctly the isotopic dependence of
energies and transition probabilities and they are in a reasonable
agreement with the available experimental data. It is worth to
mention that the anharmonicity effects are strong for the light Te
isotopes and the QRPA is not very good in such a case. The
$B(E2)$-value in the neutron-rich isotope $^{136}$Te is only
slightly larger than that of $^{134}$Te, in contrast to the trend
of Ce, Ba and Xe isotopes \cite{Rad02,Rad05,Ram01}. Such a
behaviour of $B(E2)$ is related with the shell structure in this
region and an interplay between the QRPA amplitudes for neutrons
and protons in Te isotopes.
\begin{figure}[t!]
\includegraphics[width=7cm]{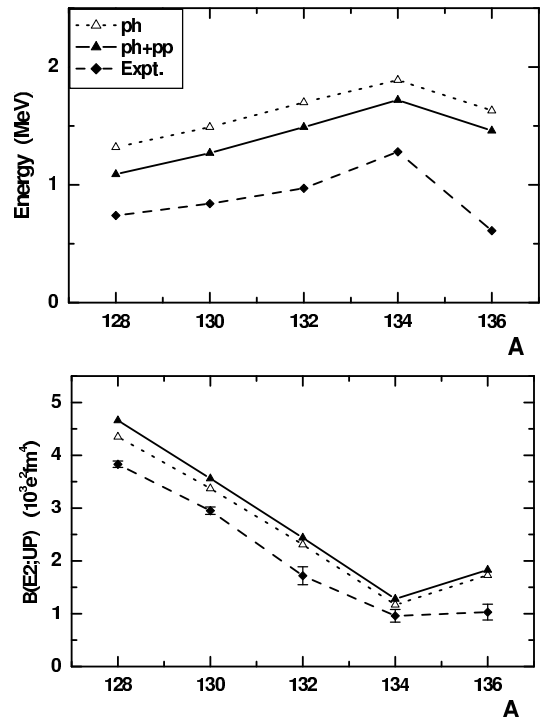}
\caption{Energies and B(E2)-values for up-transitions
to the first $2^{+}$ states in $^{128-136}$Te.}
\end{figure}
%
%
%
\subsection{N=82 isotones}
\begin{figure}[t!]
\includegraphics[width=7cm]{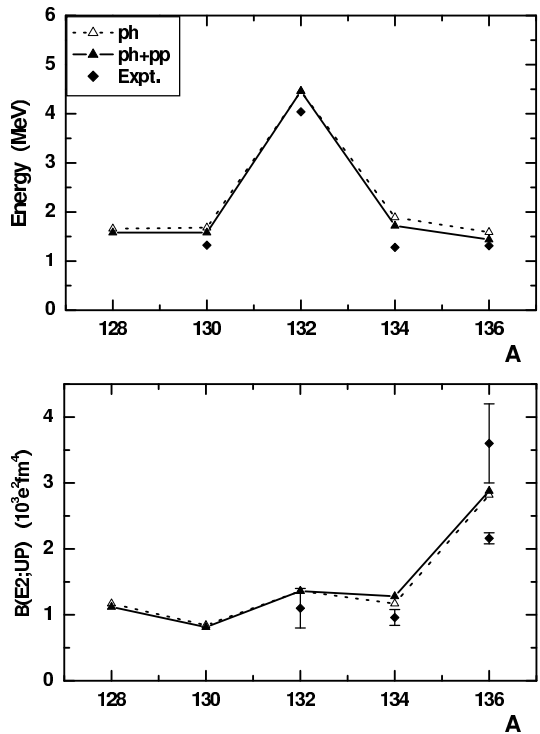}
\caption{Energies and B(E2)-values for up-transitions
to the first $2^{+}$ states in the N=82 isotones.}
\end{figure}
\begin{figure}[t!]
\includegraphics[width=6cm]{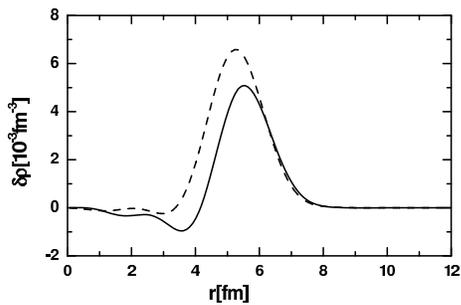}
\caption{Neutron (solid line) and proton (dashed line)
transition densities of the $2_1^+$ state of $^{130}$Cd.}
\end{figure}
It is interesting to study a change of the structure of the
$2_1^+$ states along the N=82 isotones chain. The N=82 isotones
below the doubly magic nucleus $^{132}$Sn are crucial for stellar
nucleosynthesis \cite{Jun07}. Results of our calculations and
existing experimental data \cite{Ram01,Rad05,Jak02,Jun07} are
shown in Fig.3. It is seen that the inclusion of the p-p channel
does not change energies and transition probabilities along this
chain. Going along the N=82 isotones chain one can find that
$2_1^{+}$ states in $^{128}$Pd and $^{130}$Cd have a non
collective structure with a domination of the proton configuration
$\{1g_{9/2},1g_{9/2}\}$. In $^{132}$Sn as it was discussed above
the main configurations are the neutron $\{2f_{7/2},1h_{11/2}\}$
(61\%) and the proton $\{2d_{5/2},1g_{9/2}\}$ (33\%) ones. In
$^{134}$Te and $^{136}$Xe the $2_1^{+}$ states are very collective
and many proton configurations contribute in their structure. The
structure pecularities are reflected in the B(E2) behaviour in
this chain. Higher collectivity results in an increase of the
transition probability. An additional information  about the
structure of the first $2^+$ states can be extracted from the
proton scattering experiments (for example, see \cite{Jew99}) by
looking at the ratio of the multipole transition matrix elements
$M_n/M_p$ that depends on the relative contributions of the proton
and neutron configurations. Results of our calculations are given
in Table II, where the $M_n/M_p$ ratio for $^{128}$Pd, $^{130}$Cd
is less than half of $N/Z$ value, indicating a very strong proton
contribution. According to our calculations there is a sharp
increase of $M_n/M_p$ at Z=50, N=82. Such a behaviour of the
multipole transition matrix elements $M_n/M_p$ in other nuclei can
indicate a shell closure.
\begin{table}
\caption{ $(M_n/M_p)/(N/Z)$ ratios for the first $2_1^{+}$ states}
\begin{tabular}{cccccc}
\hline\noalign{\smallskip}
Nucleus    & $^{128}$Pd & $^{130}$Cd & $^{132}$Sn    & $^{134}$Te & $^{136}$Xe\\
\noalign{\smallskip}\hline\noalign{\smallskip}
Theory     &    0.47    &  0.49      &  0.81         & 0.54       & 0.55      \\
\noalign{\smallskip}\hline
\end{tabular}
\end{table}
Another quantity that characterizes the $2_1^+$ state is the
transition density. As an example the neutron and proton
transition densities of the $2_1^+$ state of $^{130}$Cd are
displayed in Fig.~4. The neutron transition density is shifted
outwards as compared to the proton transition density due to the
presence of the neutron skin. We get the similar tendency in the
case of the other isotones but this effect becomes weak in
$^{136}$Xe.
%
%
\section{Conclusions}
A finite rank separable approximation for the QRPA calculations
with Skyrme interactions that was proposed in our previous work is
extended to take into account the residual particle-particle
interaction. This approximation enables one to reduce considerably
the dimensions of the matrices that must be inverted to perform
structure calculations in very large configuration spaces. As an
illustration of the method we have studied the energies and
transition probabilities of the $2^+_1$ states around the
$^{132}$Sn region. Using the same set of parameters we describe
available experimental data and give predictions for the N=82
isotones that are important for stellar nucleosynthesis. Including
the quadrupole p-p interaction results in a reduction of the
collectivity and this may be more important for nuclei far from
closed shells. Such calculations which take into account the
two-phonon terms in wave functions are in progress now.
%
%
\section*{Acknowledgments}
We are grateful to Prof. Ch.Stoyanov for valuable discussions.
A.P.S. and V.V.V. thank the hospitality of IPN-Orsay where a part
of this work was done. This work is partly supported by the
IN2P3-JINR agreement.
%
%
\begin{appendix}
\section {}
In this appendix, we derive the formulas which help us to represent
the antisymmetrized p-p matrix elements in the separable form in
the angular coordinates.

In Eq.(10) the sum over $\lambda$ can be trasformed into:
\begin{widetext}
\begin{eqnarray}
\sum_{\lambda }(-1)^{j_{2}+j_{3}+J+\lambda }\left\{
\begin{array}{ccc}
j_{4} & j_{3} & J \\
j_{1} & j_{2} & \lambda
\end{array}
\right\} \left\langle j_{1}\left\| i^{\lambda }Y_{\lambda }\right\|
j_{3}\right\rangle \left\langle j_{2}\left\| i^{\lambda }Y_{\lambda
}\right\| j_{4}\right\rangle  = {\hat j_{1}}{\hat j_{2}}{\hat j_{3}
}{\hat j_{4}}\left( 16\pi \right)^{-1}
i^{l_{3}+l_{4}-l_{1}-l_{2}} \nonumber \\
\times \left( \left( 1+(-1)^{l_{1}+l_{2}+l_{3}+l_{4}}\right) \left(
\begin{array}{ccc}
j_{3} & j_{4} & J \\
-\frac{1}{2} & -\frac{1}{2} & 1
\end{array}
\right) \left(
\begin{array}{ccc}
j_{1} & j_{2} & J \\
-\frac{1}{2} & -\frac{1}{2} & 1
\end{array}
\right) \right.   \nonumber \\
\left. -\left( (-1)^{l_{1}+l_{3}}+(-1)^{l_{2}+l_{4}}\right)
(-1)^{j_{1}+j_{3}}\left(
\begin{array}{ccc}
j_{3} & J & j_{4} \\
-\frac{1}{2} & 0 & \frac{1}{2}
\end{array}
\right) \left(
\begin{array}{ccc}
j_{1} & J & j_{2} \\
-\frac{1}{2} & 0 & \frac{1}{2}
\end{array}
\right) \right)
\end{eqnarray}
Then, the antisymmetrized p-p matrix elements take the form:
\begin{equation}
V_{1234}^{pp}-V_{1243}^{pp}={\hat J}^{-2}\sum_{JM}\langle
j_{1}m_{1}j_{2}m_{2}\mid JM\rangle \langle j_{3}m_{3}j_{4}m_{4}\mid
JM\rangle \left\langle j_{1}\left\| i^{J}Y_{J}\right\| j_{2}\right\rangle
\left\langle j_{3}\left\| i^{J}Y_{J}\right\| j_{4}\right\rangle I^{pp}\left(
j_{1}j_{2}j_{3}j_{4}\right)\eta_{j_{1}}\eta_{j_{2}}\eta_{j_{3}}\eta_{j_{4}}
.
\end{equation}
\end{widetext}
\section {}
Taking into account the residual p-p interaction we show how the
finite rank separable form of the residual force (\ref{mult}) can
simplify the solution of the QRPA equations (\ref{QRPA}). In the
$6N$-dimensional space we introduce a vector $\left(
\begin{array}{c}
{\cal D}_0\left( \tau \right) \\
{\cal D}_+\left( \tau \right) \\
{\cal D}_-\left( \tau \right) \\
\end{array}
\right)$
by its components:
\begin{equation}
\mathcal{D}_{\beta }^{k}\left( \tau \right) =\left(
\begin{array}{c}
D_{\beta }^{k}\left( \tau \right)  \\
D_{\beta }^{k}\left( -\tau \right)
\end{array}
\right) ,\beta =\left\{ 0,+,-\right\}
\end{equation}
\begin{equation*}
D_{0}^{\lambda ik}\left( \tau \right) =\left. \sum_{jj'}\right.
^{\tau }f_{jj'}^{(\lambda k)}u_{jj'}^{\left(
+\right) }\left( X_{jj'}^{\lambda i}+Y_{jj'}^{\lambda i}\right)
,
\end{equation*}
\begin{equation*}
D_{\pm }^{\lambda ik}\left( \tau \right) =\left. \sum_{jj'}\right.
^{\tau }f_{jj'}^{( \lambda k)}\eta_{jj'}v_{jj'}^{\left( \pm
\right) }\left( X_{jj'}^{\lambda i}\mp
Y_{jj'}^{\lambda i}\right)
.
\end{equation*}
The index k runs over the $N$-dimensional space (k=1,2,...,N).
Following our previous paper~\cite{ssvg02} the QRPA equations
(\ref{QRPA}) can be reduced to the set of equations:
\begin{widetext}
\begin{equation}
\left(
\begin{array}{ccc}
\mathcal{M}_{00}\left( \tau \right) -1 & \mathcal{M}_{0+}\left( \tau \right)
& \mathcal{M}_{0-}\left( \tau \right)  \\
\mathcal{M}_{+0}\left( \tau \right)  & \mathcal{M}_{++}\left( \tau \right) -1
& \mathcal{M}_{+-}\left( \tau \right)  \\
\mathcal{M}_{-0}\left( \tau \right)  & \mathcal{M}_{-+}\left( \tau \right)
& \mathcal{M}_{--}\left( \tau \right) -1
\end{array}
\right) \left(
\begin{array}{c}
\mathcal{D}_{0}\left( \tau \right)  \\
\mathcal{D}_{+}\left( \tau \right)  \\
\mathcal{D}_{-}\left( \tau \right)
\end{array}
\right) =0
,
\end{equation}
where ${\cal M}$ is the $2N\times 2N$ matrix
\begin{equation}
\label{M}
{\cal M}_{\beta\beta'}^{kk'}\left( \tau \right) =\left(
\begin{array}{cc}
(\kappa _0^{\left( \beta', k'\right) }+\kappa _1^{\left(
\beta', k'\right) })T_{\beta\beta'}^{kk'}\left( \tau
\right)  & (\kappa _0^{\left( \beta', k'\right) }-\kappa
_1^{\left( \beta', k'\right) })T_{\beta\beta'}^{kk'}
\left( \tau \right)  \\
(\kappa _0^{\left( \beta', k'\right) }-\kappa _1^{\left(
\beta', k'\right) })T_{\beta\beta'}^{kk'}\left( -\tau
\right)  & (\kappa _0^{\left( \beta', k'\right) }+\kappa
_1^{\left( \beta', k'\right) })T_{\beta\beta'}^{kk'}
\left( -\tau \right)
\end{array}
\right)
,
1 \leq k, k' \leq N
\end{equation}
\end{widetext}
In the definition~(\ref{M}), $\kappa^{\left( 0,
k'\right)}=\kappa^{\left( ph, k'\right)}$, $\kappa^{\left( \pm,
k'\right)}=\kappa^{\left( pp, k'\right)}$. The matrix elements
$T^{kk'}$ have the following form:
\[
T_{00}^{kk'}\left( \tau \right) =\left. \sum_{jj'}\right.
^{\tau }\chi _{jj'}^{\lambda kk'}\left(
u_{jj'}^{\left( +\right) }\right) ^{2}\varepsilon _{jj'}
,
\]
\[
T_{++}^{kk^{\prime }}\left( \tau \right) =\left. \sum_{jj'}\right.
^{\tau }\chi _{jj'}^{\lambda kk'}\left(
v_{jj'}^{(+)}\eta_{jj'}\right) ^{2}\varepsilon _{jj'}
,
\]
\[
T_{--}^{kk'}\left( \tau \right) =\left. \sum_{jj'}
\right. ^{\tau }\chi _{jj'}^{\lambda kk'}\left(
v_{jj'}^{(-)}\eta_{jj'}\right) ^{2}\varepsilon _{jj'}
,
\]
\[
T_{0+}^{kk'}\left( \tau \right)=
T_{+0}^{kk'}\left( \tau \right) =\left. \sum_{jj'}
\right. ^{\tau }\chi _{jj'}^{\lambda kk'}
u_{jj'}^{\left( +\right) }v_{jj'}^{(+)}
\eta_{jj'}\omega _{\lambda i}
,
\]
\[
T_{0-}^{kk'}\left( \tau \right) =
T_{-0}^{kk'}\left( \tau \right) =\left. \sum_{jj'}
\right. ^{\tau }\chi _{jj'}^{\lambda kk'}
u_{jj'}^{(+)}v_{jj'}^{(-)}\eta_{jj'}\varepsilon _{jj'}
,
\]
\[
T_{+-}^{kk'}\left( \tau \right) =
T_{-+}^{kk'}\left( \tau \right) =\left. \sum_{jj'}
\right. ^{\tau }\chi _{jj'}^{\lambda kk'}
v_{jj'}^{(+)}v_{jj'}^{(-)}
\left(\eta_{jj'}\right)^{2}
\omega _{\lambda i}
,
\]
where
\[
\chi _{jj'}^{\lambda kk'}=\frac{f_{jj'}^{(\lambda k)}
f_{jj'}^{(\lambda k')}}{\hat{\lambda}
^{2}\left( \varepsilon _{jj'}^{2}-\omega _{\lambda i}^{2}\right)}
.
\]
One can see that the matrix dimensions never exceed $6N \times 6N$
independently of the size of the two-quasiparticle configuration.
The excitation energies $\omega _{\lambda i}$ are the roots of the
secular equation:
\begin{equation}
\det \left(
\begin{array}{ccc}
\mathcal{M}_{00}\left( \tau \right) -1 & \mathcal{M}_{0+}\left( \tau \right)
& \mathcal{M}_{0-}\left( \tau \right)  \\
\mathcal{M}_{+0}\left( \tau \right)  & \mathcal{M}_{++}\left( \tau \right) -1
& \mathcal{M}_{+-}\left( \tau \right)  \\
\mathcal{M}_{-0}\left( \tau \right)  & \mathcal{M}_{-+}\left( \tau \right)
& \mathcal{M}_{--}\left( \tau \right) -1
\end{array}
\right) =0
.
\end{equation}
The phonon amplitudes corresponding to the QRPA eigenvalue $\omega _{\lambda i}$
are obtained by Eqs.(\ref{QRPA}) and  the  normalization condition (\ref{condition}).
\end{appendix}
%
%

\end{document}